\documentclass{article}
\usepackage{graphicx,color,epsfig}
\usepackage{varioref,float,amssymb,amsmath,amsfonts}
\usepackage{changebar,longtable}
\usepackage[latin1]{inputenc}
\usepackage{makeidx}
\usepackage[below]{placeins}
\usepackage{flafter}
\usepackage[small]{caption2}
\usepackage{subfigure}
\usepackage{pst-plot}

\newcommand{\bra}[1]{\left\langle{#1}\right\vert}
\newcommand{\ket}[1]{\left\vert{#1}\right\rangle}
\begin{document}

\markboth{P. Lara, R.~Portugal, S. Boettcher}
{Quantum Walks on Sierpinski Gaskets}


\title{Quantum Walks on Sierpinski Gaskets}

\author{Pedro Carlos S. Lara\footnote{LNCC/MCTI, Petr\'opolis, RJ 25651-075, Brazil}, Renato Portugal\footnote{LNCC/MCTI, Petr\'opolis, RJ 25651-075, Brazil,
portugal@lncc.br}, and Stefan Boettcher\footnote{Physics Dept., Emory University, Atlanta, GA 30322-2430, USA}
}

\maketitle

\begin{abstract}
We analyze discrete-time quantum walks on Sierpinski gaskets using a flip-flop shift operator with the Grover coin. We obtain the scaling of two important physical quantities: the mean-square displacement and the mixing time as function of the number of points.
The Sierpinski gasket is a fractal that lacks translational invariance and the results differ
from those described in the literature for ordinary lattices. We find that the
displacement varies with the initial location. Averaged over all initial locations,
our simulation obtain an exponent very similar to classical diffusion.
\end{abstract}

\section{Introduction}

Discrete-time quantum walks have been introduced by Aharanov,
Davidovich, and Zagury\cite{Aharonov} as the quantum version of
classical random walks. Quantum walks on lattices can spread out ballistically, in
contrast with the diffusive behavior of classical random walks. This
characteristic has motivated many studies pursuing quantum
algorithms that are faster than their classical
counterparts\cite{Shenvi,Ambainis}.

Discrete-time quantum walks have been investigated previously on many graphs. The most studied graph is the one-dimensional line\cite{NV00,ABNVW01,Kon02}. Quantum walks  have been analyzed on two-dimensional square  lattices\cite{Bartlett,MPA10}, and on the hypercube\cite{MPAD08}. A spatial search using the discrete-time quantum walk model has been undertaken on the Sierpinski gasket\cite{PR12}, and
on the Hanoi network of degree 3.\cite{MPB11} A quantum walk on the dual Sierpinski gasket using the continuous-time quantum walk model has been analyzed by Agliari \textit{et. al.}\cite{ABM10}

In this paper we focus our attention on discrete-time
quantum walks on the Sierpinski gasket.
We analyze the dynamics based on the standard evolution
operator $U=S\cdot(C\otimes I)$, where $S$ is the flip-flop shift
operator, $C$ is the coin, and $I$ is the
Identity operator. Throughout, we are using the Grover coin. The main physical quantities that we analyze are
the mean-square displacement in form of the standard deviation in position, the limiting probability
distribution, and the mixing time. The results are compared with
classical random walks on the Sierpinski gasket and with quantum walks on
other graphs, such as the square lattice.

The paper is organized as follows: In Section \ref{sec:ST} we derive the
evolution equation for quantum walks on Sierpinski gaskets. In
Section \ref{Sec:PhysQ} we present the numerical results for the standard deviation, the limiting probability distribution,
and the mixing time. In the last section, we present our
conclusions.

\section{Standard Quantum Walk Dynamics}\label{sec:ST}

\begin{figure}[!ht]
    \setcaptionmargin{0.15in}
\begin{center}
\begin{pspicture}(5.35686,4.6)
\psgrid[subgriddiv=1,xunit=0.707107,yunit=1,griddots=10](0,0)(8,4)
\psline[border=3pt]{*-*}(1.414214,0)(0.000000,0)
\psline[border=3pt]{*-*}(1.414214,0)(0.707107,1)
\psline[border=3pt]{*-*}(1.414214,0)(2.121320,1)
\psline[border=3pt]{*-*}(1.414214,0)(2.828427,0)
\psline[border=3pt]{*-*}(2.828427,0)(1.414214,0)
\psline[border=3pt]{*-*}(2.828427,0)(2.121320,1)
\psline[border=3pt]{*-*}(2.828427,0)(3.535534,1)
\psline[border=3pt]{*-*}(2.828427,0)(4.242641,0)
\psline[border=3pt]{*-*}(4.242641,0)(2.828427,0)
\psline[border=3pt]{*-*}(4.242641,0)(3.535534,1)
\psline[border=3pt]{*-*}(4.242641,0)(4.949747,1)
\psline[border=3pt]{*-*}(4.242641,0)(5.656854,0)
\psline[border=3pt]{*-*}(0.707107,1)(0.000000,0)
\psline[border=3pt]{*-*}(0.707107,1)(1.414214,0)
\psline[border=3pt]{*-*}(0.707107,1)(2.121320,1)
\psline[border=3pt]{*-*}(0.707107,1)(1.414214,2)
\psline[border=3pt]{*-*}(2.121320,1)(2.828427,0)
\psline[border=3pt]{*-*}(2.121320,1)(1.414214,0)
\psline[border=3pt]{*-*}(2.121320,1)(0.707107,1)
\psline[border=3pt]{*-*}(2.121320,1)(1.414214,2)
\psline[border=3pt]{*-*}(3.535534,1)(2.828427,0)
\psline[border=3pt]{*-*}(3.535534,1)(4.242641,0)
\psline[border=3pt]{*-*}(3.535534,1)(4.949747,1)
\psline[border=3pt]{*-*}(3.535534,1)(4.242641,2)
\psline[border=3pt]{*-*}(4.949747,1)(5.656854,0)
\psline[border=3pt]{*-*}(4.949747,1)(4.242641,0)
\psline[border=3pt]{*-*}(4.949747,1)(3.535534,1)
\psline[border=3pt]{*-*}(4.949747,1)(4.242641,2)
\psline[border=3pt]{*-*}(1.414214,2)(0.707107,1)
\psline[border=3pt]{*-*}(1.414214,2)(2.121320,1)
\psline[border=3pt]{*-*}(1.414214,2)(2.828427,2)
\psline[border=3pt]{*-*}(1.414214,2)(2.121320,3)
\psline[border=3pt]{*-*}(2.828427,2)(1.414214,2)
\psline[border=3pt]{*-*}(2.828427,2)(2.121320,3)
\psline[border=3pt]{*-*}(2.828427,2)(3.535534,3)
\psline[border=3pt]{*-*}(2.828427,2)(4.242641,2)
\psline[border=3pt]{*-*}(4.242641,2)(4.949747,1)
\psline[border=3pt]{*-*}(4.242641,2)(3.535534,1)
\psline[border=3pt]{*-*}(4.242641,2)(2.828427,2)
\psline[border=3pt]{*-*}(4.242641,2)(3.535534,3)
\psline[border=3pt]{*-*}(2.121320,3)(1.414214,2)
\psline[border=3pt]{*-*}(2.121320,3)(2.828427,2)
\psline[border=3pt]{*-*}(2.121320,3)(3.535534,3)
\psline[border=3pt]{*-*}(2.121320,3)(2.828427,4)
\psline[border=3pt]{*-*}(3.535534,3)(4.242641,2)
\psline[border=3pt]{*-*}(3.535534,3)(2.828427,2)
\psline[border=3pt]{*-*}(3.535534,3)(2.121320,3)
\psline[border=3pt]{*-*}(3.535534,3)(2.828427,4)
\psline[border=4pt]{->}(-0.5,0)(-0.5,2)
\psline[border=4pt]{->}(0,-0.5)(2,-0.5) \rput(-0.7,1){$y$}
\rput(1,-0.7){$x$}
\end{pspicture}

\

\end{center}
\caption{Sierpinski gasket of generation $g=2$ embedded in
a two-dimensional plane.}
\label{fig:sierpinki}
\end{figure}
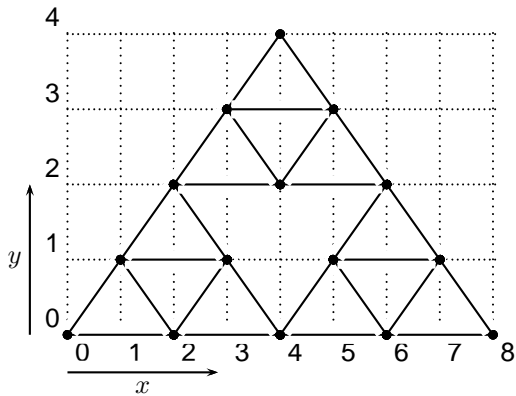

The Sierpinski gasket of generation $g$ is a degree-4 regular graph,
an example of which is depicted in Fig.~\ref{fig:sierpinki} for
$g=2$. It has $N=3(3^g+1)/2$ nodes and a
fractal or Hausdorff dimension of $d_f=\log 3/\log 2$, which is larger
than for the line and smaller than for the plane.

A coined quantum walk on the Sierpinski gasket embedded in
the two-dimensional plane has
a Hilbert space ${\cal H}_C\otimes {\cal H}_P$, where ${\cal H}_C$
is the $4$-dimensional coin subspace and ${\cal H}_P$ is the
$N$-dimensional position subspace.
${\cal H}_P$ is spanned by vectors of type $\ket{x,y}$
with integers $0 \leq x\leq 2^{g+1}$ and $0 \leq y\leq 2^g$ restricted to be on the gasket, as shown in Fig.~\ref{fig:sierpinki}.
Due to the embedding, we use
the computational basis $\{\ket{k}, 0\leq k\leq5\}$ for
the coin space ${\cal H}_C$, but only four of these basis vectors
are utilized for each vertex.

The shift operator for the internal vertices is
\begin{equation}\label{pq_S_x_y_0}
    S\ket{k}\ket{x,y}=\ket{-k}\ket{x+f(k),y+g(k)},
\end{equation}
where $-k$ is the inverse of $k$ modulo 6. The coin value is inverted
after the shift (flip-flop shift). Functions $f$ and $g$ are defined in
the Table~\ref{tab:f_g}.

\begin{table}[h!]
  \centering
  \begin{tabular}{|c|c|c|c|c|c|c|}
    \hline
     & 0 & 1 & 2 & 3 & 4 & 5 \\
    \hline
    $f$ & 2 & 1 & -1 & -2 & -1 & 1 \\
    \hline
    $g$ & 0 & 1 & 1 & 0 & -1 & -1 \\
    \hline
  \end{tabular}
  \label{tab:f_g}
  \caption{Auxiliary functions for the shift operator.}
\end{table}

For the external vertices $(0,0)$, $(2^g,2^g)$, $(2^{g+1},0)$, where
$g$ is the generation level, we consider two cases of boundary
conditions: (1) periodic and (2) reflective. In case (1), the action
of the shift operator is given
by
\begin{eqnarray*}
  S\ket{3}\ket{0,0} &=& \ket{-3}\ket{2^{g+1},0} \\
  S\ket{4}\ket{0,0} &=& \ket{-4}\ket{2^g,2^g} \\
  S\ket{1}\ket{2^g,2^g} &=& \ket{-1}\ket{0,0} \\
  S\ket{2}\ket{2^g,2^g} &=& \ket{-2}\ket{2^{g+1},0} \\
  S\ket{0}\ket{2^{g+1},0} &=& \ket{-0}\ket{0,0} \\
  S\ket{5}\ket{2^{g+1},0} &=& \ket{-5}\ket{2^g,2^g}.
\end{eqnarray*}
Those special cases can be implemented through functions $f$ and $g$
and, in this case, $f$ and $g$ will depend on the location $(x,y)$.
In case (2), the action of the shift
operator is given by
\begin{eqnarray*}
  S\ket{3}\ket{0,0} &=& \ket{4}\ket{1,1} \\
  S\ket{4}\ket{0,0} &=& \ket{3}\ket{2,0} \\
  S\ket{1}\ket{2^g,2^g} &=& \ket{1}\ket{2^g-1,2^g-1} \\
  S\ket{2}\ket{2^g,2^g} &=& \ket{2}\ket{2^g+1,2^g-1} \\
  S\ket{0}\ket{2^{g+1},0} &=& \ket{5}\ket{2^{g+1}-1,1} \\
  S\ket{5}\ket{2^{g+1},0} &=& \ket{0}\ket{2^{g+1}-2,0}.
\end{eqnarray*}

The Grover coin is defined as
\begin{equation}\label{pq_Grover_x_y}
    G = {2 \ket{\textrm{D}}\bra{\textrm{D}}-I},
\end{equation}
where $\ket{\textrm{D}}=\frac{1}{2}\sum_{k=0}^3\ket{k}$. Its {matrix
representation} is
\begin{eqnarray}
  G = \frac{1}{2}
        \begin{bmatrix}
          -1 & \,\,\,\,1 & \,\,\,\,1 & \,\,\,\,1 \\
          \,\,\,\,1 & -1 & \,\,\,\,1 & \,\,\,\,1 \\
          \,\,\,\,1 & \,\,\,\,1 & -1 & \,\,\,\,1 \\
          \,\,\,\,1 & \,\,\,\,1 & \,\,\,\,1 & -1 \\
        \end{bmatrix}.
\end{eqnarray}

The generic state of the walker at time $t$ is described by
\begin{equation}\label{pq_generic_state_x_y}
\ket{\Psi(t)}=\sum_{k=0}^{5}\sum_{{x}=0}^{2^{g+1}}\sum_{{y}=0}^{\min\{x,2^{g+1}-x\}}\psi_{k;\,x,y}(t)\ket{k}\ket{x,y},
\end{equation}
where the coefficients $\psi_{k;\,x,y}(t)$ are complex functions
that obey the normalization condition
\begin{equation}\label{pq_normaliz_x_y}
    \sum_{k=0}^{5}\sum_{{x,y}}\left| \psi_{k;\,x,y}(t) \right|^2 = 1,
\end{equation}
for all time $t$.

Applying the evolution operator
\begin{equation}\label{pq_U_hiper_x_y}
    U=S\ (G\otimes I)
\end{equation}
to the generic state, we obtain
\begin{eqnarray}
\ket{\Psi(t+1)} &=& \sum_{k,k'=0}^{5}\sum_{x,y} \psi_{k';\,{x,y}}(t) \,
G^{(x,y)}_{k,k'}\,\ket{-k}\ket{x+f(k),\,y+g(k)}.
\end{eqnarray}
Renaming the dummy indices, we obtain
\begin{eqnarray}\label{pq_generic_state_t1_x_y}
   \ket{\Psi(t+1)} &=& \sum_{k,k'=0}^{5}\sum_{x,y} G^{(x-f(-k),y-g(-k))}_{-k,\,k'}\,\psi_{k';\,{x-f(-k),\,y-g(-k)}}(t)\ket{k}\ket{{x,y}}.
\end{eqnarray}
Expanding the left hand side of the above equation in the
computational basis and equating  like coefficients, we obtain the
evolution equation for the quantum walk,
\begin{eqnarray}\label{pq_eq_evol_x_y}
\psi_{k;\,x,y}(t+1) &=& \sum_{k'=0}^{5}
G^{(x-f(-k),y-g(-k))}_{-k,\,k'}\, \psi_{k';\,{x-f(-k),\,y-g(-k)}}(t).
\end{eqnarray}
The matrix $G^{(x,y)}$ depends on $x,y$, since there are six types of
vertices that are distinct in their orientation. For each one, we have to use the correct labels for their
edges.

We use Eq.~(\ref{pq_eq_evol_x_y}) to numerically simulate the
evolution of the quantum walk using initial conditions of
the form $\ket{\textrm{D}}\ket{x,y}$,
where $\ket{\textrm{D}}=\frac{1}{2}\sum_{k=0}^3\ket{k}$ is
the uniform vector in the coin space. Note that $\ket{\textrm{D}}$
is not biased. The same is true for the Grover coin $G$. This
coin and the flip-flop shift operator play an important
role in spatial search algorithms\cite{AKR05,PR12}.

\section{Physical Quantities}\label{Sec:PhysQ}

In this section, we analyze the behavior of quantum walks
on the Sierpinski gasket with the focus on the diffusion processes.
The main physical quantities that we analyze are the mean-square displacement in form of
the standard deviation in position, and the mixing time.


\subsection{Standard Deviation}

The physical quantities that we will analyze are defined using
the probability distribution over the vertices of the graph.
It is one of the main physical quantities that is available
in the analysis of the behavior of quantum walks.
The probability distribution is given by
\begin{equation}\label{pdf}
    p(t;x,y) =  \sum_{k=0}^{5}\left| \psi_{k;\,x,y}(t) \right|^2.
\end{equation}


The position standard deviation $\sigma(t)$ is defined as
\begin{equation}\label{}
    \sigma(t)^2=\sigma_x(t)^2+\sigma_y(t)^2,
\end{equation}
where
\begin{eqnarray}
  \sigma_x(t)^2 &=& \sum_x x^2\, p(t,x) - \left(\sum_x x\, p(t,x)\right)^2, \nonumber\\
  \sigma_y(t)^2 &=& \sum_y y^2\, p(t,y) - \left(\sum_y y\, p(t,y)\right)^2, \nonumber
\end{eqnarray}
and
\begin{eqnarray*}
  p(t,x) &=& \sum_y p(t,x,y), \\
  p(t,y) &=& \sum_x p(t,x,y).
\end{eqnarray*}

For a walker that starts located on a specific vertex,
the standard deviation at intermediate times $1\ll t\ll t_{\rm co}$ increases as a power-law\begin{equation}\label{MSD}
\sigma(t)\sim a\,t^{\frac{1}{d_w}},
\end{equation}
which defines the diffusion exponent $1\le d_w<\infty$, in analogy to a classical walk. On a finite system, the walk eventually reaches the farthest vertex which cuts off the growth in $\sigma$  at some time $t_{\rm co}$, beyond that it
oscillates around an average value. In order to analyze the
diffusion process, we are interested
in the asymptotic behavior of the power law regime for the infinite system, $g\to\infty$. Therefore, on any finite Sierpinski gasket, we bound the evolution time
to be smaller than the time that the walker takes
to reach the farthest vertex.  To measure displacement (standard deviation), we use
reflective boundary conditions and the flip-flip shift
operator.

\begin{figure}[!ht]
    \setcaptionmargin{0.18in}
    \centering %
    \subfigure {\includegraphics[width=7cm,angle=0,scale=1.2]{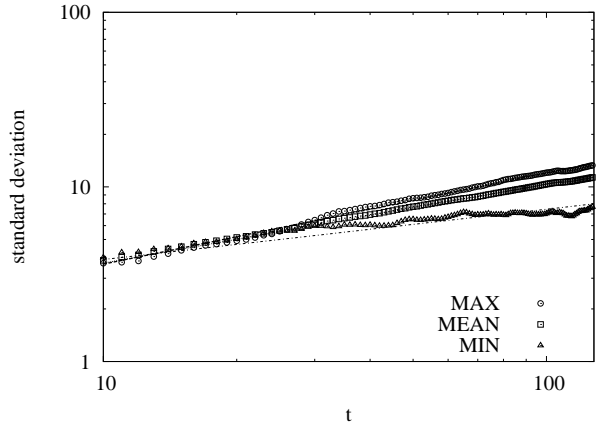}}
    \caption{{\footnotesize Standard deviation $\sigma(t)$ of a flip-flop quantum walk on the
    Sierpinski gasket $(g=8)$
      after 128 steps: (1) with initial state $\ket{\textrm{D}}\ket{247,5}$ displayed with circles, (2) with initial state $\ket{\textrm{D}}\ket{256,224}$ displayed with triangles, and (3) averaged over all initial
      states $\ket{\textrm{D}}\ket{x,y}$, $\forall (x,y)$ displayed as squares.
      }} %
    \label{fig:standev1}
\end{figure}

The first result we have obtained from the simulations,
which is strikingly different from the behavior of
quantum walks on lattices, is that the scale of the
standard deviation depends on the initial
vertex. For example, for generation $g=8$, the
fastest growth in the displacement is obtained when the
walker starts on vertex $(x=247,y=5)$
Fig.~\ref{fig:standev1} shows the standard deviation as function
of the number of steps of a quantum walk
with the initial state $\ket{\textrm{D}}\ket{247,5}$.
Our fit yields $\sigma=1.1t^{0.52}$, i.e., $d_w\approx1.92$, significantly faster than classical diffusion ($d_w=\log_{2}5=2.32\ldots$) but still considerably slower than the standard deviation for quantum
walks on a square lattice, for which $d_w=1$.

The slowest-growing standard deviation of a flip-flop quantum walk
on the Sierpinski gasket of generation $g=8$ is obtained
for the initial state $\ket{\textrm{D}}\ket{256,224}$.
The best fit in this case using the data in Fig.~\ref{fig:standev1} is $\sigma\sim 2.0t^{0.29}$ (or $d_w\approx3.45$),
which shows a very small spreading rate characterizing a
sub-diffusive process.


\begin{figure}[!ht]
    \setcaptionmargin{0.15in}
    \centering %
    \includegraphics[width=6cm,angle=0,scale=1.2]{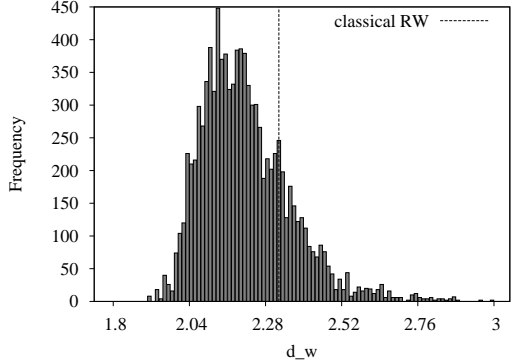}
    \caption{\footnotesize Histogram of the frequency of the initial
    conditions that have
    the same scale in the standard deviation of a flip-flop
    quantum walk on the Sierpinski gasket at $g=8$
    with the initial state
    $\ket{\textrm{D}}\ket{x,y}$, $\forall (x,y)$. A vertical line indicates the corresponding result for classical diffusion.
      } %
    \label{fig:histogram}
\end{figure}

Since the displacement apparently depends on the initial vertex,
it is interesting to define a mean standard
deviation $\bar\sigma(t)$ as function
of time in the following way
\begin{equation}\label{}
    \bar\sigma(t)=\frac{1}{N}\sum_{x,y} \sigma_{x,y}(t),
\end{equation}
where sub-indices $x,y$ of $\sigma_{x,y}(t)$ indicate the
initial location used to obtain the standard deviation. From each location the walker evolves
from an initially uniform state in coin space.
Fig.~\ref{fig:standev1} depicts the behavior of $\bar\sigma(t)$.
The numerical results suggest a
best fit of $\bar\sigma(t)\sim1.3t^{0.44}$, or $d_w\approx2.27$.
Average over all initial locations makes this scaling exponent $d_w$ a characteristic of the Sierpinski
gasket of generation $g=8$, which happens to be remarkably close to the result for classical diffusion, $d_w=\log_{2}5=2.32\ldots$.
The histogram in Fig.~\ref{fig:histogram} shows
the number of such initial conditions that have the same fitted exponent, $d_w$, in
Eq.~(\ref{MSD}). For example, for $d_w=2.14$ there
are around 450 vertices that can be used as initial condition
to obtain the same scaling.  The range in scaling (for $g=8$) extends from $d_w\approx1.92$ to 3.45, although the bulk of the distribution is centered very close to the result for classical diffusion, marked by a vertical line. The width of this distribution is only about 17\% of the mean, and it would be interesting to see whether the width narrows further for increasing system sizes $g\to\infty$.

\subsection{Limiting Distribution}

The average probability distribution is
given by
\begin{equation}\label{tm_barP_v_t}
    {\bar p}(T,x,y)=\frac{1}{T} \sum_{t=0}^{T-1} p(t,x,y).
\end{equation}
Note that ${\bar p}(T,x,y)$ is a probability distribution for all
$T$, because $$\sum_{x,y=0}^{N-1} {\bar p}(T,x,y)=1.$$

The interpretation of ${\bar p}(T,x,y)$ uses {projective
measurements},\index{projective measurement} therefore ${\bar
p}(T,x,y)$ evolves {stochastically}, and
converges to a limiting distribution when $T$ goes to
infinity. The definition of the limiting probability distribution is
\begin{equation}
    \pi(x,y) = \lim_{T\rightarrow\infty} {\bar p}(T,x,y).
\end{equation}
This limit exists and can be calculated explicitly if the
expressions for the eigenvalues of the evolution operator are known. The limiting
distribution depends on the initial condition in general.

\begin{figure}[!ht]
    \setcaptionmargin{0.13in}
    \centering %
     \subfigure {\includegraphics[width=7cm,angle=0,scale=1]{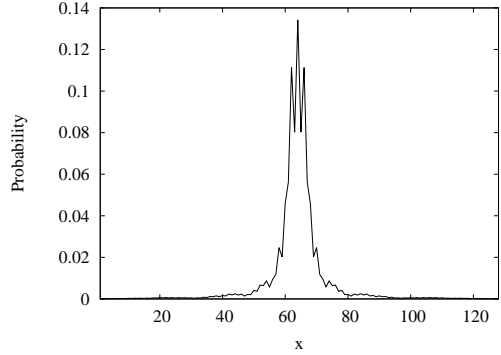}} 
    \caption{{\footnotesize Limiting distribution of a flip-flop quantum walk on the Sierpinski gasket
    $(g=6)$ with the initial
      state $\ket{\textrm{D}}\ket{64,0}$ as function of position $x$. We have
      added all probabilities with different values of $y$ having
      the same value of $x$.
      The boundary conditions are periodic.
      }} %
    \label{fig:limitingdist1}
\end{figure}

Fig.~\ref{fig:limitingdist1} shows the limiting distribution as a function
of position $x$ of a
flip-flop quantum walk that departs from the central bottom vertex $(x=256,y=0)$
with periodic boundary conditions.
The probabilities in $y$-direction have been added up to 
generate a one-dimensional plot.
In general, the numerical simulations show that
the limiting distribution for the Sierpinski
gasket depends on the initial condition and
is highly concentrated around the initial vertex. Note that
a walker encounters frequent bottlenecks that inhibit spreading.
If the walker starts in vertex $(256,0)$,
there are only two passage points, $(128,128)$ and
$(384,128)$, towards the top of the Sierpinski gasket (in
Fig.~\ref{fig:sierpinki} these points correspond to $(2,2)$ and $(6,2)$).
A three-dimensional plot shows that the limiting probability is very close to zero
when $y>1$.

\begin{figure}[!ht]
    \setcaptionmargin{0.15in}
    \centering %
    \subfigure {\includegraphics[width=7cm,angle=0,scale=1]{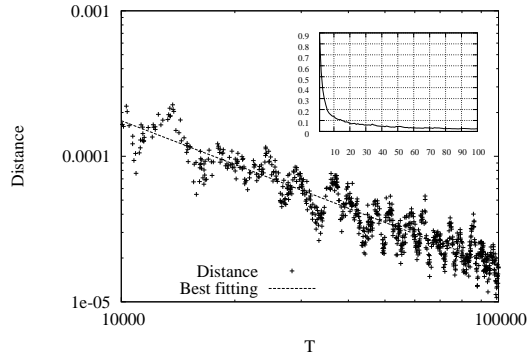}} 
    \caption{\footnotesize Total variation distance between ${\bar p}(T,x,y)$ and $\pi(x,y)$ for a flip-flop quantum walk on the Sierpinski gasket $(g=6)$ with the initial
      state $\ket{\textrm{D}}\ket{64,0}$.
      The best fit is $\|p(T,x,y)-\pi(x,y)\|=1.76/T$. 
      The inset show the plot without using the log-log scale.
      } %
    \label{fig:dtv}
\end{figure}

The average distribution ${\bar p}(T,x,y)$ converges to the limiting distribution $\pi(x,y)$.
This can be confirmed by the graph of the distance between these distributions as function
of time.  The total variation distance between two
probability distributions $p(x,y)$ and $q(x,y)$ is defined as
\begin{equation}\label{rm_D_p_q}
    \parallel p-q \parallel =\frac{1}{2} \sum_{x,y} \left|p(x,y)-q(x,y)\right|.
\end{equation}
Fig.~\ref{fig:dtv} shows the plot of $\|\bar p(T,x,y)-\pi(x,y)\|$ as function of the
number of steps represented by $T$. The best fit suggests that this
distance scales as $1/T$. 
We note that it decays approximately as $1/T$ not only for the Sierpinski
gasket but also for two-dimensional lattices\cite{MPA10} and hypercubes\cite{MPAD08}. 
This numerical result can be explained analytically. Expanding the
quantum walk state in the eigenbasis that diagonalizes the evolution operator,
one can obtain an explicit expression for $ \|\bar p(T,x,y)-\pi(x,y)\|$ for 
any initial condition. The expression has the form
\begin{equation}
    \|\bar p(T,x,y)-\pi(x,y)\| = \frac{\sum_{x,y}\left|\sum_c c_{x,y} \left({{\textrm e}^{2\pi i \Delta_c T } - 1}\right)\right|}{T},
\end{equation}
where $\Delta_c$ is a difference between two non-equal eigenvalues, $c_{x,y}$ is a constant, and the inner sum is over all pairs of non-equal eigenvalues. Some results of Aharonov \text{et. al.}\cite{AAKV01} help to obtain that analytical expression. The modulus of the term ${{\textrm e}^{2\pi i \Delta_c T } - 1}$ in the numerator of $ \|\bar p(T,x,y)-\pi(x,y)\|$ is a bounded oscillatory function. So, the distance between the average and the limiting distribution scales as $1/T$ in general and oscillates around the curve $1/T$, confirming the data shown in Fig.~\ref{fig:dtv}.

\subsection{Mixing Time}

The {mixing time} $\tau_\epsilon$ is defined as
\begin{equation}
    \tau_\epsilon = \min\big\{T\,|\,\forall t\ge T,\,\parallel{\bar p}(t,x,y)-\pi(x,y)\parallel\le \epsilon\big\},
\end{equation}
which can be interpreted as the smallest number of steps such that
the distance between the average distribution and the limiting
distribution becomes permanently smaller than $\epsilon$. 
If $\|\bar p(T,x,y)-\pi(x,y)\|$ obeys an inverse power law as a function of time, then $\tau_{\epsilon}$ obeys an inverse power law as a function of $\epsilon$.

\begin{figure}[!ht]
    \setcaptionmargin{0.15in}
    \centering %
    \includegraphics[width=7cm,angle=0,scale=1]{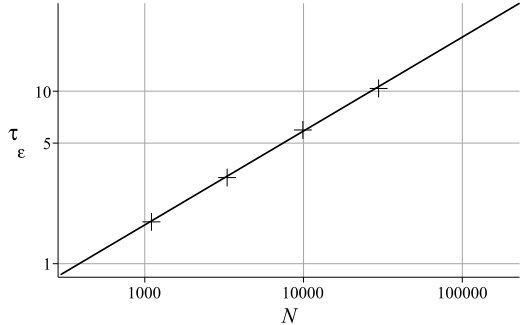}
    \caption{\footnotesize The mixing time of a flip-flop quantum walk on the Sierpinski gasket with the initial
      state $\ket{\textrm{D}}\ket{2^g,0}$.
      The best fit is $\tau_\epsilon=0.034N^{0.54}$. The first point corresponds to $g=6$.} %
    \label{fig:mixingtime}
\end{figure}

The mixing time depends on
the initial condition in general and on the size $N$ of the graph. We have
generated the same kind of data of Fig.~\ref{fig:dtv} for Sierpinski gasket of
generation 6 up to 10. Using the best-fitting curves we can estimate $\tau_\epsilon$
as a function of $N$.
Fig.~\ref{fig:mixingtime} shows that
$\tau_\epsilon$ has a power law in terms of the number of vertices. The data allows us
to estimate that
\begin{equation}
    \tau_\epsilon = O\left(\frac{N^{0.54}}{\epsilon}\right)
\end{equation}
when we take $\ket{\textrm{D}}\ket{2^g,0}$ as initial state.

\section{Conclusions}

We have analyzed the flip-flop discrete-time quantum walk on the Sierpinski
gasket of finite generation embedded in the two-dimensional plane
using reflective and periodic boundary conditions. Our investigations focus on the
following physical quantities: the position standard deviation 
(with reflective boundary conditions) and
the mixing time (with periodic boundary conditions). 
Performing numerical simulations on Sierpinski gaskets up
to generation $g=10$, we have obtained the scaling exponent of the standard
deviation as function of the number of steps and the scale of the
mixing time as function of the number of vertices.

For the system sizes studied, the results depend significantly on the initial condition. As fractal lattices lack translational invariance, quantum interference effects likely vary strongly with the initial location. A characteristic
 way to assign a distinct diffusion exponent $d_w$ to the Sierpinski gasket
is provided by performing an average over all  initial locations.
In this case, we obtain an average exponent $\bar{d}_w\approx2.27$ that is remarkably close to the result for classical diffusion, $d_w=\log_{2}5=2.32\ldots$, on this system.
Therefore, a quantum walk on the Sierpinski gasket spreads slower than
 on a square lattices.


The limiting distribution for the Sierpinski gasket depends on the
initial condition and is concentrated around the initial vertex.
This happens with other graphs such as the two-dimensional lattice\cite{MPA10}.
The scaling of the mixing time for the Sierpinski gasket $O(N^{0.54}/\epsilon)$
is close to the scaling of  mixing time for the two-dimensional lattice, which
is believed\cite{MPA10} to be $O(\sqrt{N\log N}/\epsilon)$. Our data is not precise enough
to determine the presence of a term that depends on $\log N$. The result differs 
from the scaling on the cycle which is believed\cite{AAKV01} to be  $O(N\log N/\epsilon)$.


\section*{Acknowledgments}
We acknowledge financial support from CNPq. SB is grateful for the support and hospitality of LNCC during this project.

\end{document}